\documentclass[fleqn,usenatbib]{mnras}

\usepackage{newtxtext,newtxmath}

\usepackage[T1]{fontenc}

\DeclareRobustCommand{\VAN}[3]{#2}
\let\VANthebibliography\thebibliography
\def\thebibliography{\DeclareRobustCommand{\VAN}[3]{##3}\VANthebibliography}

\usepackage{graphicx}	
\usepackage{amsmath}	
\usepackage{amssymb}	
\usepackage{graphicx}
\usepackage{bm}
\usepackage{color}

\title[Constraints on Alfv\'enic Turbulence]{Constraining Alfv\'enic Turbulence with Helicity Invariants}

\author[Mahajan \& Lingam]{
Swadesh M. Mahajan,$^{1}$\thanks{E-mail: mahajan@mail.utexas.edu}
and Manasvi Lingam$^{2}$\thanks{E-mail: mlingam@fit.edu}
\\
$^{1}$Institute for Fusion Studies, The University of Texas at Austin, Austin, TX 78712\\
$^{2}$Department of Aerospace, Physics and Space Sciences, Florida Institute of Technology, Melbourne FL 32901, USA\\
}
\date{Accepted XXX. Received YYY; in original form ZZZ}

\pubyear{2020}

\begin{document}
\label{firstpage}
\pagerange{\pageref{firstpage}--\pageref{lastpage}}
\maketitle

\begin{abstract}
In this paper, we study the constraints imposed by the invariants (generalized helicities and energy) of extended magnetohydrodynamics on some global characteristics of turbulence. We show that the global turbulent kinetic and magnetic energies will approach equipartition only under certain circumstances that depend on the ratio of the generalized helicities. In systems with minimal thermal energy, we demonstrate that the three invariants collectively determine the characteristic length scale associated with Alfv\'enic turbulence.
\end{abstract}

\begin{keywords}
(Sun:) solar wind -- turbulence -- plasmas --  magnetic fields
\end{keywords}



\section{Introduction}
The existence of (generalized) magnetic helicity constraints introduces a fundamental distinction between the Navier-Stokes fluid turbulence and the low-frequency Alfv\'enic turbulence realized in magnetized plasma systems. The resultant Alfv\'enic turbulence has been widely investigated within the context of the simplest model - namely, ideal magnetohydrodynamics (MHD) - but also for models that are often collectively known as ``beyond MHD'' or extended MHD \citep{GP04,Frei14}.

The scientific literature is replete with examples wherein helicity invariants have been exploited to find new relaxed states. The most famous among them are the so-called Woltjer-Taylor states of ideal MHD (${\bf\nabla\times B}= \mu {\bf B}$) that are obtained by minimizing the magnetic energy $\langle|{\bf B}|^2\rangle$ while holding the magnetic helicity $h_m=\langle{\bf A}\cdot{\bf B}\rangle$ fixed \citep{Wol1,Tay1,Berg99,MM12}; henceforth, we shall make use of the notation $\langle \dots \rangle = \int d^3x$ for the sake of simplicity.

A very crucial role played by the constancy of $h_m$ in the evolution of MHD turbulence was identified in early MHD simulations as well as in analytical models: it permitted the inverse cascading of magnetic helicity in 3D models \citep{FPLM75,PFL76},\footnote{However, at scales smaller than the electron skin depth, the inverse cascade of helicity is transformed into a direct cascade as per theory and simulations \citep{MLM17,MMT18}.} whereas in standard fluid turbulence, the transfer of energy and helicity is typically from larger to smaller scales \citep{Moff78,KR80,Bis03,ZMD04,BS05,Gal18}. The latter feature was inherent in the famous conjecture of Andrey Nikolaevich Kolmogorov \citep{Kol41} that led to the equally famous scaling law $E_k\sim k^{-5/3}$ for the kinetic energy spectrum $E_k$ \citep{Fri,Bis03,Dav}. Another seminal result in the realm of MHD turbulence is the Iroshnikov-Kraichnan theory \citep{Iro63,Kra65}, which modelled turbulent fluctuations as weakly interacting Alfv\'enic wave packets and yielded the magnetic energy spectrum $E_k\sim k^{-3/2}$ \citep{Bis03}.

This paper, although motivated by Kolmogorov's legacy, will dwell on precisely those features of Alfv\'enic turbulence that are absent in Navier-Stokes systems. The goal, in the spirit of \citet{Kol41}, is to obtain results of maximal simplicity and, hopefully, of considerable generality that are potentially valid for all Alfv\'enic turbulence irrespective of its origination and evolution. More precisely, we will delve into constraints on Alfv\'enic turbulence imposed by the helicity and energy invariants of extended MHD, and thereby extend prior analyses along similar lines \citep{OSYM,OSM,MNSY}; see also \citet{Hel17}. 

As we shall show henceforth, this line of enquiry yields several results of broad scope and interest: (1) Total turbulent energy in each channel - namely, magnetic ($E_m$), kinetic ($E_\mathrm{kin}$) and thermal ($E_\mathrm{th}$) - is determined by a single attribute of turbulence, namely, a characteristic length scale ($L_T=K_T^{-1}$), (2) Complete expressions in terms of a single unknown parameter for all these energies in terms of the invariants and $L_T$, thus enabling us to predict, for instance, the relative energy distribution.

\section{Invariants of extended MHD}
For the sake of simplicity, we will concentrate on a two component (electron-ion) quasineutral plasma. Under the assumption of isotropic pressure with an adiabatic equation of state $(p \propto n^{\gamma})$, each component obeys the following equation of motion \citep{SI,SM1,ML1}:
\begin{equation}\label{eqm1}
{\frac{\partial}{\partial t}}\, {\bf P}_\beta = {\bf v}_\beta\times{\bf\Omega}_\beta-\nabla{\psi_\beta},
\end{equation}
where ${\bf P}_\beta={\bf A} + (m_\beta c/q_\beta){\bf v}_\beta$ is proportional to the canonical momentum,
${\bf\Omega}_\beta = {\bf\nabla\times\bf P}_\beta = {\bf B}+(m_\beta c/q_\beta)\nabla\times{\bf v}_\beta$ represents the generalized vorticity for the species $\beta$ with mass and charge of $m_\beta $ and $q_\beta$, and ${\psi_\beta}= c/q_\beta(h_\beta  + 1/2 m_\beta v_\beta^2+ q_\beta \phi)$ encompasses all of the gradient forces; note that $h_\beta$ is the specific enthalpy, and $\phi$ is the electrostatic potential.

Taking the curl of Eq.~(\ref{eqm1}) yields the canonical vortical dynamics \citep{MY98,SM1,AGMD} that is epitomized by
\begin{equation}
\frac{\partial{\bf\Omega}_\beta}{\partial t}=\nabla\times\left({\bf v}_\beta\times
{\bf\Omega}_\beta\right), \label{eqmotion2}
\end{equation}
The low frequency behavior of this system of ideal fluid equations, which is closed via the Amp\`ere's law, 
\begin{equation} \label{Amplaw}
{\bf\nabla\times\bf B}=(4\pi/c){\bf J}, \quad\quad {\bf J}=\sum q_\beta n_\beta{\bf v}_\beta,
\end{equation}
is the object of this investigation. Straightforward manipulation of (\ref{eqm1})-(\ref{Amplaw}) yields the following three constants of motion: the total energy 
\begin{equation}
E=\Bigg\langle{\frac{B^2}{2}}+{\frac{1}{2}}\sum_\beta n_\beta m_\beta v^2_\beta+{\frac{p}{(\gamma-1)}}\Bigg\rangle,\label{EnCons}
\end{equation}
where $p$ is the total pressure, and two generalized helicities (GH), 
\begin{equation}
H_\beta=\frac{1}{2}\langle {\bf P}_\beta\cdot{\bf\Omega}_\beta\rangle,
\label{HelicCons}
\end{equation}
associated with each species. For a perfectly conducting system of $n$ dynamical species, there exist a total of $(n+1)$ bilinear invariants \citep{ML1}. Although it is self-evident, it must nevertheless be emphasized that in any magneto-fluid system, unless the fluid inertia is neglected, it is the generalized helicity $H_\beta$, and \emph{not} the magnetic helicity $H_m= \langle{{\bf A}\cdot{\bf\ B}}\rangle$  that is conserved; for instance, the conservation of $H_m$ in MHD and in Hall MHD holds true because electron inertia is ignored \citep{Turn86}.

To study the constrained Alfv\'en dynamics (including turbulence), it is convenient to work in the equivalent one fluid variables, viz., 
the center-of-mass velocity ${\bf V}$, and the current ${\bf J}$ defined below:
\begin{equation} 
{\bf V}=\frac{m{\bf v_e}+ M{\bf v_i}}{m+M}= \mu_e {\bf v_e}+ \mu_i {\bf v_i}, \quad  {\bf J}= ne({\bf v_i}-{\bf v_e})
\label{1fluvariables}
\end{equation}
where the two species are identified as electrons (mass $m$ and charge $-e$) and protons (mass $M$ and charge $e$). However, we shall not rewrite 
(\ref{eqm1})-(\ref{eqmotion2}) explicitly in terms of ${\bf V}$ and ${\bf J}$ because, in what follows, we will focus only on the invariants
(\ref{EnCons})-(\ref{HelicCons}). The electron and ion helicity invariants 
translate into the new variables as
\begin{eqnarray}\label{elec-ion helicities} 
2H_e = \left<{\bf \hat A}\cdot {\bf \hat B} + (m/e)^2 {\bf V}\cdot{\bf\nabla\times\bf V} -2 (m/e){\bf V}\cdot{\bf \hat B}\right> \\
2H_i = \left<{\bf \hat A}\cdot {\bf \hat B} + (M/e)^2 {\bf V}\cdot{\bf\nabla\times\bf V} +2 (M/e){\bf V}\cdot{\bf \hat B}\right>
\end{eqnarray}
where ${\bf \hat A}= {\bf A}+\lambda_e^2 {\bf\nabla\times\bf B}$ is the vector potential modified by the contribution stemming from a finite electron skin depth $(\lambda_e^2=c^2/{\omega_{pe}}^2)$; in other words, the second term in ${\bf \hat A}$ is obtained after using the Amp\`ere's law given by (\ref{Amplaw}); it is also derivable by means of the Hamiltonian or Lagrangian formulations from the parent two-fluid model \citep{KC14,AKY15,LMM15,LMM16,DML16}.

Evidently, both helicities comprise of their purely magnetic (${\bf \hat A}\cdot {\bf \hat B}$), purely kinematic  (${\bf V}\cdot{\bf\nabla\times\bf V}$), and the mixed (i.e., cross) (${\bf V}\cdot{\bf \hat B}$) components - the chief difference is that the contribution of the kinematic and mixed parts can be far more dominant for the protons (due to $M\gg m$). For further analysis, it is much more transparent to construct the invariant combinations:
\begin{equation}\label{plus- helicity} 
 H_+= 2\mu_{i}H_i+2\mu_{e}H_e= \left<{\bf \hat A}\cdot {\bf \hat B}\right> + \frac{m}{M} \lambda_i^2 \left<{\bf V}\cdot{\bf\nabla\times\bf V}\right> 
\end{equation}
\begin{equation}\label{minus- helicity} 
H_-= 2H_i - 2H_e= \lambda_i^2 \left<{\bf V}\cdot{\bf\nabla\times\bf V}\right> +2 \lambda_i \left<{\bf V}\cdot{\bf \hat B}\right> 
\end{equation}
where the magnetic field has been normalized to some ambient field strength $B_0$ and and the velocity field is measured in terms of the corresponding Alfv\'en speed $V_A$ (where $V_A^2= B_0^2/(4\pi n M)$), i.e., we have used Alfv\'enic units \citep{ML1}. Notice that, aside from the normalized fields, the only basic parameter is the ion skin depth $\lambda_i$ (where $\lambda_i^2 = c^2/{\omega_{pi}}^2)$ that defines the intrinsic length scale of the system. Of course, the existence of the term proportional to $(m/M)$ serves as a reminder that the electron inertia is not (yet) neglected and the electron length scale ($\lambda_e= {\sqrt {(m/M)}} \lambda_i$) appears in $H_+$ and in the variables ${\bf \hat A}$ and ${\bf \hat B}$. With the above choice of normalization, the helicities acquire the dimensions of length. In what follows, we shall utilize the dimensionless helicities defined to be $h_\pm=H_\pm/\lambda_i$.

These invariants $h_+$ and $h_-$ act in concordance to constrain the total magnetic and kinetic energies of the system. We suppose that the system is embedded in an ambient magnetic field such that ${\bf \hat B} = {\bf \hat B_0} + {\bf \hat b}$, and a similar expression can be constructed for the vector potential ${\bf \hat A}$.\footnote{We implicitly presume that the functions ${\bf \hat B}$ and ${\bf \hat A}$ are well-behaved and that the term $\left<{\bf \hat A_0}\cdot {\bf \hat B_0}\right>$ is finite.} For the time being, we analyze the case where there exists no ambient flow, implying that ${\bf v}$ fully represents the velocity field (i.e., we have ${\bf V}$=${\bf v}$). Note that $\left<{\bf \hat A}\cdot {\bf \hat B}\right>$ will acquire a contribution of the form $\left<{\bf \hat A_0}\cdot {\bf \hat B_0}\right>=H_0$, while the linear terms will vanish on integration \citep{KR80}. Written fully in terms of the (normalized) fluctuating fields denoted by lowercase boldface letters, our normalized invariant equations become
\begin{equation} \label{plus-turb helicity} 
h = h_+ - h_0=  \frac{\left<{\bf \hat a}\cdot {\bf \hat b}\right>}{\lambda_i} + \frac{m}{M} \lambda_i \left<{\bf v}\cdot{\bf\nabla\times\bf v}\right>, 
\end{equation}
\begin{equation}\label{minus-turb helicity} 
h_-= \lambda_i \left<{\bf v}\cdot{\bf\nabla\times\bf v}\right> + 2\left<{\bf v}\cdot{\bf \hat b}\right> 
\end{equation}

\section{Helicity constraints on turbulence}
From this point onward, our analysis will be purely algebraic and qualitative, as it relies essentially on heuristic considerations. In this paper, we will neglect the electron scale length ($\lambda_e=0$) for the sake of simplicity, although electron inertia can be readily reintroduced; in other words, we investigate the Hall MHD regime \citep{GP04}. In this scenario, the second term on the RHS of (\ref{plus-turb helicity}) becomes vanishingly small, and we end up with ${\bf \hat a}= {\bf a}$ and ${\bf \hat b}= {\bf b}$. We wish to figure out the constraints imposed on Alfv\'enic turbulence by the invariance of $h$ and $h_-$. 

\begin{figure}
\includegraphics[width=7.5cm]{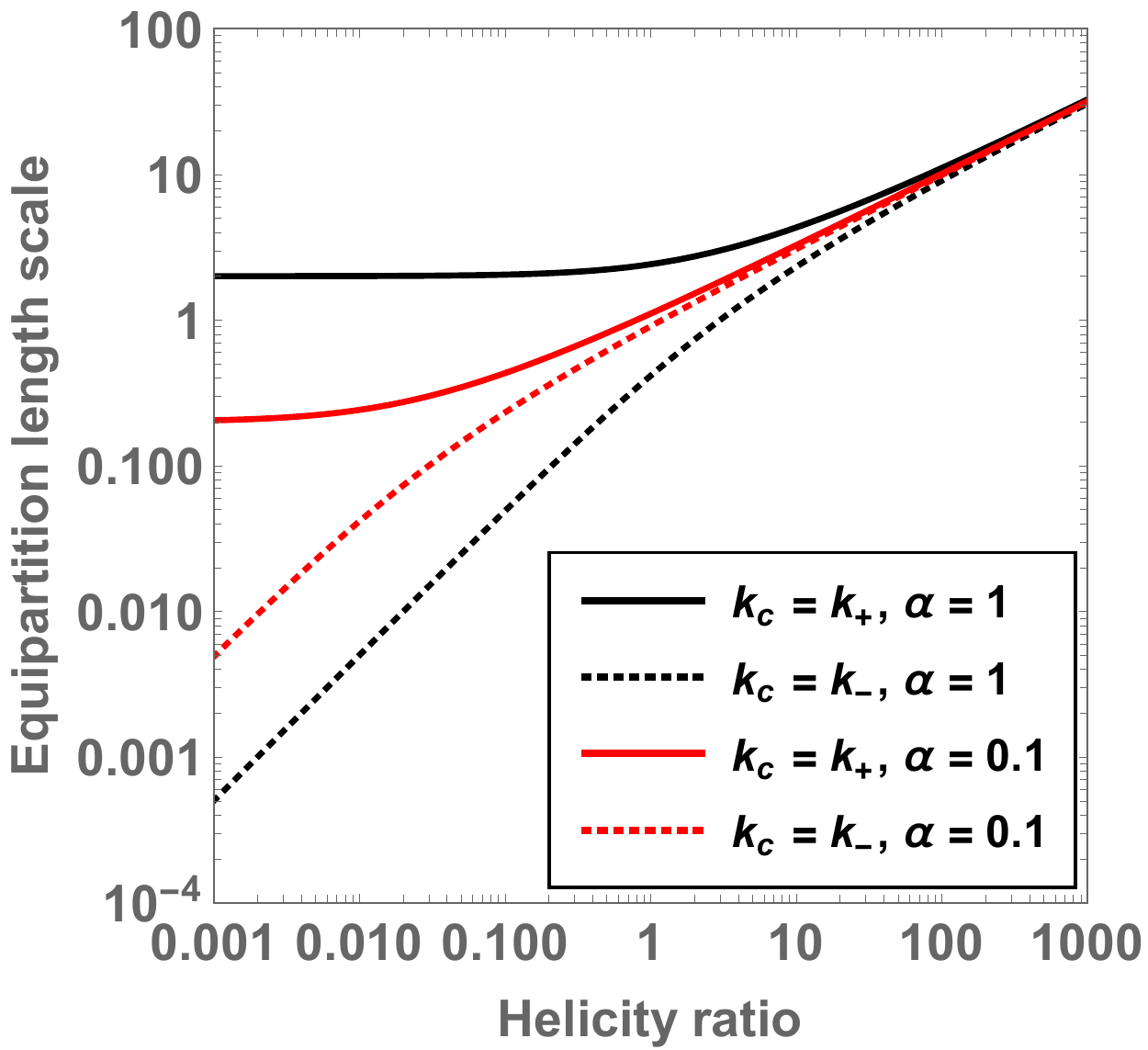} \\
\caption{The equipartition length scale ($k_c$) as a function of the helicity ratio $h_-/h$. There are two different solutions for $k_c$, namely, $k_+$ and $k_-$ depending on whether $s_+$ or $s_-$ is adopted. We have plotted $k_+$ and $k_-$ for two choices of $\alpha$, viz. $\alpha = 1$ and $\alpha = 0.1$.}
\label{FigEqScale}
\end{figure}

Now, we introduce a characteristic length scale $L_T$ for the turbulent magnetic field ${\bf b}$; the equivalent wave number is $K_T=1/L_T$. More specifically, because ${\bf b}=\nabla\times{\bf a}$ is valid, we will invoke a phenomenological scaling of the form ${\bf b} \sim K_T {\bf a}$ or ${\bf a} \sim K_T^{-1}{\bf b}$. In other words, one may interpret $K_T$ as the measure of the gradient associated with ${\bf b}$; a similar approach was introduced for the turbulent velocity in \citet[pg. 348]{PFL76}. Note, however, that this mathematical expression is valid \emph{sensu stricto} if ${\bf b}$ is specified to be a Beltrami field, with $K_T$ serving as the corresponding Beltrami parameter; this ansatz is not unreasonable because a number of publications model the turbulent fields as Arn'old-Beltrami-Childress fields \citep{CD95,BS05}. By utilizing the relationship ${\bf a}= K_T^{-1}{\bf b}$ introduced above, (\ref{plus-turb helicity}) reduces to
\begin{equation}\label{H} 
h\approx \frac{\langle{\bf b}\cdot{\bf b}\rangle}{k_T}
\end{equation}
where $k_T= \lambda_i  K_T$ is the inverse of turbulent scale length measured in units of the ion skin depth. Rewriting (\ref{H})
yields an estimate for the magnetic energy
\begin{equation}\label{Em} 
E_m=\frac{\langle{\bf b}\cdot{\bf b}\rangle}{2}\approx \frac{h k_T}{2}
\end{equation}
Following the same procedure we obtain 
\begin{equation}\label{Hminus} 
h_-\approx 2k_{T} E_\mathrm{kin} +4\alpha \sqrt{E_\mathrm{kin}}\sqrt {E_m}
\end{equation}
where $E_\mathrm{kin}=\left<{\bf v}\cdot{\bf v}\right>/2$. The second term on the RHS represents an alignment condition of sorts, because we suppose that the dimensionless factor $\alpha$ captures the ``projection'' of one turbulent field on the other. This approach is inspired by the fact that, in a special class of exact solutions of nonlinear Alfv\'en waves, ${\bf b}$ is linearly proportional to ${\bf v}$; see, for instance, \citet{Wal44,MK05,MM09,ALM16}. A more general and rigorous strategy for obtaining this term relies upon invoking the well-known Cauchy-Bunyakovsky-Schwarz inequality \citep{JMS}, which yields
\begin{equation}
|\left<{\bf b}\cdot{\bf v}\right>|^2 \leq \left<{\bf b}\cdot{\bf b}\right> \left<{\bf v}\cdot{\bf v}\right>    
\end{equation}
and subsequently replacing the inequality in this expression with an equality involving the phenomenological dimensionless factor $\alpha$ that implicitly obeys $0 \leq \alpha \leq 1$ as follows:
\begin{equation}
|\left<{\bf b}\cdot{\bf v}\right>|^2 = \alpha^2 \left<{\bf b}\cdot{\bf b}\right> \left<{\bf v}\cdot{\bf v}\right>    
\end{equation}
Lastly, we make use of the definitions of $E_\mathrm{kin}$ and $E_m$ introduced earlier, and take the square root of the above equation to obtain the second term on the RHS of (\ref{Hminus}).

By utilizing (\ref{Em}), (\ref{Hminus}) is readily solved for 
\begin{equation}\label{Ekin} 
E_\mathrm{kin} \approx \frac{s_\pm^{2}}{k_T}, \quad s_\pm=\frac{-\alpha \sqrt{h} \pm \sqrt{\alpha^2 h + h_-}}{\sqrt{2}}
\end{equation}
where $s$ depends on the constants of motion and the parameter $\alpha$; it will be regulated by the detailed nature of turbulence. The estimates for the global turbulent magnetic and kinetic energies, i.e., (\ref{Em}) and (\ref{Ekin}), are rather robust for all Alfv\'enic turbulence accessible within the two-fluid equations and constitutes one of the salient results in the paper. It can be readily verified that the dominant behavior, contained in the scaling, 
\begin{equation}\label{Emag} 
E_m \propto {k_T},   \quad\quad\quad E_\mathrm{kin} \propto \frac{1}{k_T}
\end{equation}
holds true (with some corrections on the order of $k_T \lambda_e$) even when the electron dynamics is retained. 

Independent of details, Alfv\'enic turbulence is strongly constrained by the ideal invariants of the system. For instance, these systems must obey a definitive, verifiable proportionality emerging from (\ref{Emag}):
\begin{equation}\label{Eratio} 
\frac{E_m}{E_\mathrm{kin}} \propto k_T^2
\end{equation}
Therefore, the short-scale turbulence ought to be much richer in magnetic energy while the portion of kinetic energy increases (in relative terms) as one moves toward longer scales (see \citealt[Fig. 3]{SS09}). To put it differently, from (\ref{Eratio}) we see that the two energies may be displaced from equipartition. This lack of equipartition is commonly observed in studies of Hall MHD turbulence, dynamos, and ``reverse'' dynamos \citep{MGM,KM04a,MSMS,MAP07,LM15,LB,LB16}. In particular, the above behavior is consistent with numerical simulations of magnetic and kinetic energy spectra, as seen from \citet[Fig. 2]{MA14} and \citet[Fig. 4]{SP15}; note, however, that the plots in these publications investigate energy spectra and not the global energy budgets. Furthermore, a number of MHD turbulence simulations \citep{WBP11,OMWO} as well as observations of the (turbulent) solar wind \citep{MG82,GVM91,SMBV,CBSM} have revealed an ``excess'' of magnetic energy at small scales as well as differences in the slopes of magnetic and kinetic spectra \citep{BPBP}; the theoretical calculations by \citet{ALM16} suggest that this feature is a generic characteristic of extended MHD. As the invariants are the defining ``labels'' for a given system, once they are specified, we can determine explicit estimates for both ${E_m}$ and $E_\mathrm{kin}$.  

\begin{figure}
\includegraphics[width=7.5cm]{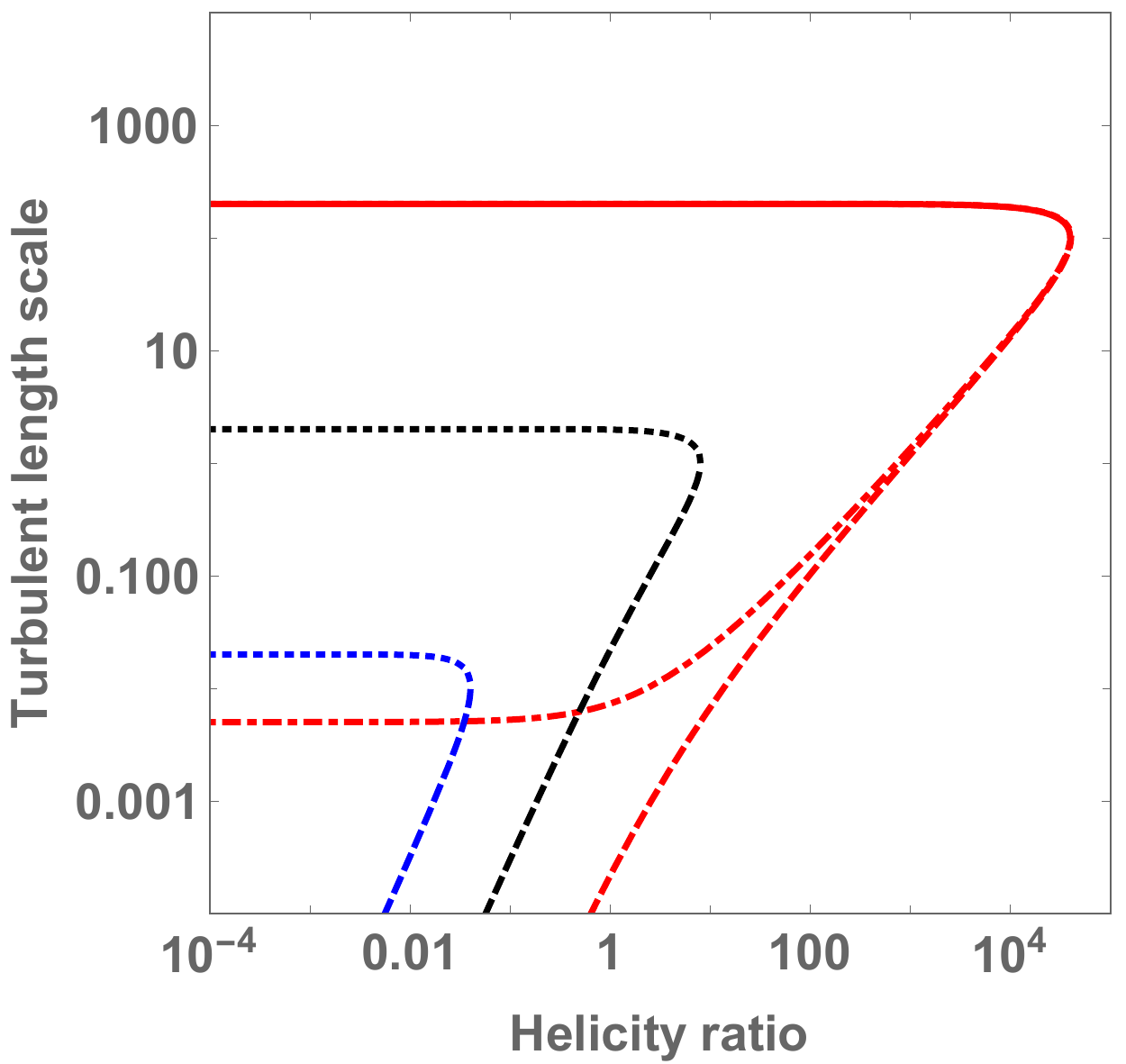} \\
\caption{The turbulent length scale is shown for different solutions and values of $h_-/h$ and $E/h$. The red, black and blue curves correspond to $E/h = 100$, $E/h =1$ and $E/h = 0.01$, respectively. The unbroken and dotted curves correspond to selecting the positive branch of (\ref{ScLeSimp}) with $k_c = k_+$ and $k_c = k_-$ respectively, whereas the dot-dashed and dashed curves constitute the positive branch of (\ref{ScLeSimp}) with $k_c = k_+$ and $k_c = k_-$ respectively. In all cases depicted herein, we have adopted $\alpha = 1$ for simplicity.}
\label{FiTurbScale}
\end{figure}

Notice that, although the estimate for ${E_m}$ is rather simply related to $h$ (which is essentially the magnetic helicity), $E_\mathrm{kin}$ has two solutions $s_\pm$. It is straightforward to verify that it is $s_+$ that must correspond to conventional MHD turbulence - in the limit $h_-\ll h$, 
\begin{equation}\label{Splus} 
s_+ \simeq \frac{h_-}{2\sqrt{2}\alpha \sqrt{h}}
\end{equation}
The complete expression for $s_+$ is the relevant expression for conventional Hall MHD. The larger root $s_-$ in terms of magnitude will consequently yield a higher kinetic energy. Thus, it must be emphasized that, for a given set of helicities, there are two distinct turbulent energy states: $[E_m, \,E_\mathrm{kin}(s_+)]$ and  $[E_m, \,E_\mathrm{kin}(s_-)]$. The ratio $E_m/E_\mathrm{kin}$ is physically relevant since it represents the ratio of the magnetic and kinetic energies. It is possible for this ratio to attain values both greater and smaller than unity. The critical turbulent length scale ($k_c$) at which equipartition is obtained is found by solving for $E_m/E_\mathrm{kin} = 1$, thus leading us to
\begin{equation}\label{kcfin}
    k_c = \frac{\sqrt{2} |s_\pm|}{|\sqrt{h}|} = |\alpha| \left(\sqrt{1 + \frac{h_-}{\alpha^2 h}} \pm 1\right).
\end{equation}
As expected, there are two different critical length scales at which equipartition of kinetic and magnetic energies is achieved. An interesting point that emerges from the above formula is that $k_c$ depends only on the ratio $\Gamma = h_-/h$ and not the individual helicities; aside from this ratio, it also depends on $\alpha$. For the case with $\Gamma \gg 1$, we determine that both roots converge to $k_c \approx \sqrt{h_-/h}$. Note, however, that this solution is physically problematic because it corresponds to $L_T \ll \lambda_i$ - in this regime, Hall MHD is not accurate because electron inertia effects (neglected herein) come into play. 

On the other hand, when we consider $\Gamma \ll 1$, we find that two divergent values for $k_c$ follow - we obtain $k_{c1} \approx 2$ for one branch and $k_{c2} \approx h_-/(2|\alpha| h)$ in the other. For the first branch, we arrive at $L_{c1} \approx \lambda_i/2$, whereas the second branch yields $L_{c2} \gg \lambda_i$. Thus, for $\Gamma \ll 1$, there is a manifest bifurcation of the equipartition length scales: one of them is comparable to the ion skin depth, while the other is much larger than $\lambda_i$, and presumably comparable to the characteristic system length scale \citep{YMO}. The combination of $\Gamma \ll 1$ and $L_{c2} \gg \lambda_i$ essentially means that the system is dominated by the magnetic helicity (as opposed to the cross helicity and fluid helicity) and that equipartition is being achieved at macroscopic scales \emph{sensu lato}. Hence, this regime is consistent with an ideal MHD-like picture, wherein large-scale behavior and magnetic helicity are dominant. The different values of $k_c$ as a function of $h_-/h$ and $\alpha$ are depicted in Fig. \ref{FigEqScale}.

\begin{figure}
\includegraphics[width=7.5cm]{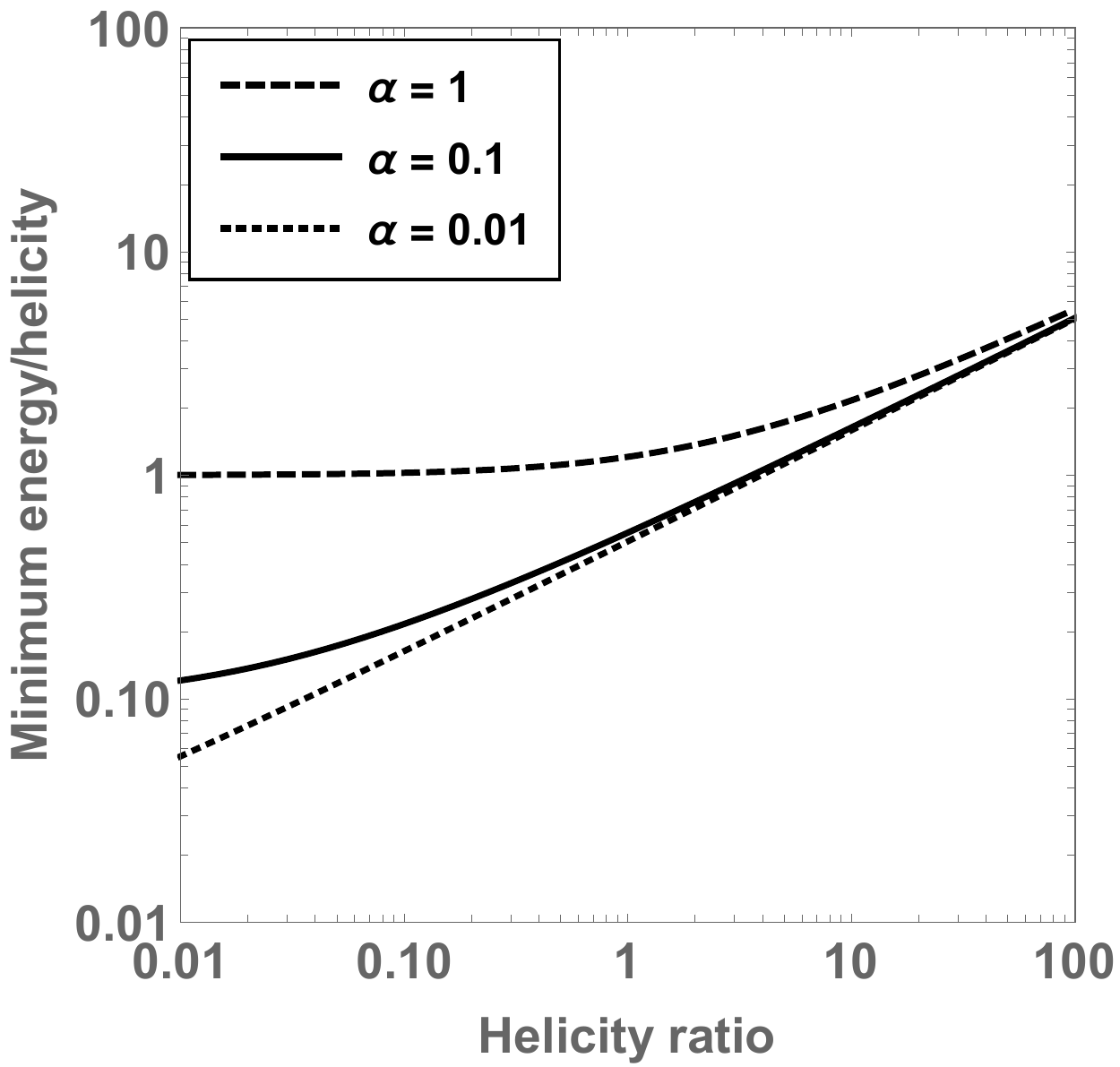} \\
\caption{The minimum value of $E/h$ that is sufficient to yield real values of $k_T$ is plotted as a function of the helicity ratio $h_-/h$ for different choices of $\alpha$. }
\label{FigMinEn}
\end{figure}

Until now, our analysis has concentrated only on the constraints on the magnetic and kinetic energy imposed by the helicity invariants. Let us now examine these results in conjunction with the conservation of energy, which in the language of preceding considerations, becomes (after having subtracted ambient field energy) 
\begin{equation}\label{Energy} 
E = \left<\frac{ {\bf b}\cdot{\bf b} +{\bf v}\cdot{\bf v}}{2}+\frac{p}{\gamma-1}\right> = E_m+E_\mathrm{kin}+E_\mathrm{th},
\end{equation}
where $E$ is a constant that denotes the difference between the total energy and ambient magnetic energy. The obvious inference is that having already estimated ${E_m}$ and $E_\mathrm{kin}$ in (\ref{Em}) and (\ref{Ekin}), we find that (\ref{Energy}) allows us to calculate the turbulent thermal energy in terms of the three invariants ($h$, $h_-$ and $E$) of the system.

If we specialize to the special case where the turbulent kinetic energy is negligible for an incompressible plasma, we see that (\ref{Energy}) reduces to
\begin{equation}\label{Energydef} 
E= E_m+E_\mathrm{kin},
\end{equation}
and thereby imposes an additional constraint on the system. In fact, we end up constraining the characteristic scale length of turbulence as follows:
\begin{equation}\label{Scalelength} 
k_T = \lambda_i K_T=\frac{\lambda_i}{L_T}= \frac{E}{h}\pm\sqrt{ \frac{E^2}{h^2}-\frac{s_\pm^2}{2h}}. 
\end{equation}
It is far more transparent to rewrite (\ref{Scalelength}) in terms of $k_c$ because we end up with
\begin{equation}\label{ScLeSimp}
  k_T =  \frac{E}{h} \left[1 \pm \sqrt{1 - \left(\frac{k_c h}{2 E}\right)^2}\right]
\end{equation}
Although this expression looks deceptively simple, it is quite complex. It has a dependence on $\alpha$, $E/h$ and $h_-/h$ via $k_c$. Moreover, there are four solutions in total: $2$ arising from the $\pm$ in the right-hand-side of (\ref{ScLeSimp}) and $2$ more from the fact that $k_c$ has two different branches as seen from (\ref{kcfin}). After fixing $\alpha$, we have plotted $k_T$ in Fig. \ref{FiTurbScale}. Note that not all of the $4$ solutions are guaranteed to be real, as seen from inspecting this figure.

In order for $k_T$ to be real-valued, the following inequality must hold true:
\begin{equation}\label{InEq}
    \Big|\frac{k_c h}{2 E}\Big| \leq 1,
\end{equation}
which imposes a constraint on the parameter space of $\{E,\,h,\,h_-\}$. Hence, depending on the parameters adopted, it is possible for $k_T$ to have either $4$ real roots, $2$ real roots and $1$ complex-conjugate pair, or $0$ real roots and $2$ complex-conjugate pairs (see Fig. \ref{FiTurbScale} for an example). These results, as embodied by the equations (\ref{Scalelength}) and (\ref{ScLeSimp}), are qualitatively similar, albeit derived from a more generic standpoint, to the generalized magneto-Bernoulli mechanism elucidated in \citet{OSYM,MNSY,SMB19}; it has been proposed that this mechanism may constitute a viable explanation for solar flares \citep{KM10}, as opposed to classic paradigms such as fast magnetic reconnection \citep{Bis00,SM11,CLH16}.

Note that (\ref{InEq}) provides us with another means of envisioning $k_c$. It is not only the length scale at which the ratio of magnetic and kinetic energies equals unity, but also the length scale that fulfills the criterion $L_{c1} \geq \lambda_i |h/E|/2$; here, recall that $L_{c1}$ is constructed from the larger root of $k_c$, which we had dubbed $k_{c1}$. By utilizing (\ref{InEq}), we have plotted the minimum value of $E/h$ that suffices to ensure that $k_T$ is real-valued; this lower bound depends on both the helicity ratio $h_-/h$ and $\alpha$.

Upon inspecting (\ref{Scalelength}), we find that $k_T$ is determined almost wholly in terms of the three constants of motion. We notice that there are two different solutions for $k_T$ for a given choice of $s_\pm$, thereby giving rise to multiple length scales that can differ considerably in magnitude. In the above setting, it would seem then that the turbulence in each incompressible Alfv\'enic system, which is defined by its three invariants, should give rise to characteristic length scales that are fully determined or severely constrained by the invariants. 

\section{Discussion}
It makes intuitive sense that integral invariants (helicities and energy) would consequently set constraints on the global (i.e., integral) magnetic, kinetic and thermal energies; in fact, they may even formally determine them. Thus, we have obtained explicit relationships between $E_m, E_\mathrm{kin}$ and $E_\mathrm{th}$ on the one hand and $L_T$ on the other, but the essentially heuristic arguments developed in this paper cannot give any specifications for the parameter $\alpha$, which measures the degree of alignment of the two turbulent fields. 

Likewise, it is natural to contend that our analysis would not directly yield the $k$-spectrum of $E_m, E_\mathrm{kin}$ and $E_\mathrm{th}$. This information can seemingly emerge only via detailed studies of the Alfv\'enic dynamics, which has been a most active field of investigation in the physics of turbulent plasmas. Thus, to reiterate, our work does not examine the consequences for energy spectra, as it focuses on the \emph{global} energy budgets. The details of the energy spectrum for the various regimes of extended MHD have been explored by \citet{ALM16} in the context of the solar wind. It was shown therein that the spectrum in the MHD regime obeys a Kolmogorov scaling as opposed to the Iroshnikov-Kraichnan scaling, in agreement with prior theoretical and empirical results \citep{GRM,BC16,SHH20}.

By simply harnessing the fundamental plasma invariants (helicities and energy), we were able to formulate certain interesting and possibly generic results for Alfv\'enic turbulence. For a specific set of invariants that remain invariant during whatever dynamics the system undergoes (thus serving as a ``label''), all three components of the total turbulent energy are potentially dictated by a single feature of turbulence that embodies the length scale $L_T$ associated with the small-scale magnetic field; conversely, one may interpret this length scale as being fully determined if the trio of invariants are specified, as seen from (\ref{ScLeSimp}). 

Expressing the turbulent energies in terms of $L_T$ also enabled us to deduce some basic constraints on their magnitudes. Although our analysis was expressly concerned with global quantities, we found that our results are compatible with spectral relationships that have been identified in the Hall regime via numerical simulations; while this fact does not validate our predictions, it bolsters their credibility. Apart from establishing some fundamental intrinsic features of Alfv\'enic turbulence as described hitherto, our results can motivate as well as provide a check on detailed simulations.

Lastly, our analysis is valid over a broad range of turbulent scale lengths, namely, ${L_{eq}}^{-1}\ll k_T= \lambda_i K_T \ll M/m$, where $L_{eq}$ is some equilibrium scale length that typically encapsulates the system size. It can be readily extended to and beyond the electron skin depth ($\lambda_e$), but it seems relatively unlikely that the characteristic scale for Alfv\'enic turbulence would enter this regime. 

\section*{Acknowledgements}
We thank our reviewer, Mitchell Berger, for the positive and insightful report. This work was partially supported by the US-DOE grant DE-FG02-04ER54742.


\bsp	
\label{lastpage}
\end{document}